\documentstyle[12pt]{article}     
\def\beq{\begin{equation}}
\def\eeq{\end{equation}}
\def\gweak{SU(2)_W}

\def\ket#1{\vert #1\rangle}
\def\leapprox{\lower.6ex
              \hbox{\kern+.2em$\buildrel <\over\sim$\kern+.2em}}
\def\cle{c_L^e}
\def\clnue{c_L^{\nu_e}}
\def\clmu{c_L^\mu}
\def\clnumu{c_L^{\nu_\mu}}

\def\clesq{\left(c_L^e\right)^2}
\def\clnuesq{\left(c_L^{\nu_e}\right)^2}

\def\slesq{\left(s_L^e\right)^2}
\def\slnuesq{\left(s_L^{\nu_e}\right)^2}
\def\slmusq{\left(s_L^\mu\right)^2}
\def\slnumusq{\left(s_L^{\nu_\mu}\right)^2}
\def\sresq{\left(s_R^e\right)^2}
\def\srnuesq{\left(s_R^{\nu_e}\right)^2}
\def\srmusq{\left(s_R^\mu\right)^2}

\def\cllisq{\left(c_L^{\ell_i}\right)^2}
\def\clnuisq{\left(c_L^{\nu_i}\right)^2}
\def\sllisq{\left(s_L^{\ell_i}\right)^2}
\def\slnuisq{\left(s_L^{\nu_i}\right)^2}
\def\srlisq{\left(s_R^{\ell_i}\right)^2}
\def\srnuisq{\left(s_R^{\nu_i}\right)^2}
\def\clu{c_L^u}
\def\cli{c_L^i}
\def\sru{s_R^u}
\def\sri{s_R^i}
\def\clc{c_L^c}
\def\src{s_R^c}
\def\cls{c_L^s}
\def\srs{s_R^s}
\def\slusq{\left(s_L^u\right)^2}
\def\sldsq{\left(s_L^d\right)^2}
\def\slssq{\left(s_L^s\right)^2}
\def\slcsq{\left(s_L^c\right)^2}
\def\clisq{\left(c_L^i\right)^2}
\def\srisq{\left(s_R^i\right)^2}

\def\sltausq{\left(s_L^\tau\right)^2}
\def\srtausq{\left(s_R^\tau\right)^2}
\def\srusq{\left(s_R^u\right)^2}
\def\srdsq{\left(s_R^d\right)^2}
\def\srssq{\left(s_R^s\right)^2}
\def\srcsq{\left(s_R^c\right)^2}
\def\slbsq{\left(s_L^b\right)^2}
\def\srbsq{\left(s_R^b\right)^2}
\def\slnutausq{\left(s_L^{\nu_\tau}\right)^2}
\def\srqsq{\left(s_R^q\right)^2}

\def\npb#1#2#3{{\it Nucl.\ Phys.}\ {\bf B#1}, (19#2) #3}
\def\plb#1#2#3{{\it Phys.\ Lett.}\ {\bf #1B}, (19#2) #3}
\def\prd#1#2#3{{\it Phys.\ Rev.}\ {\bf D#1}, (19#2) #3}
\def\prl#1#2#3{{\it Phys.\ Rev.\ Lett.}\ {\bf #1}, (19#2) #3}
\def\zpc#1#2#3{{\it Zeit.\ Phys.}\ {\bf C#1}, (19#2) #3}
\def\etal{{\it et.\ al.}}

\begin{document}
\begin{flushright}
UdeM-LPN-TH131\\
December, 1992
\end{flushright}
\null\vskip 1.5truecm
\centerline{\Large\bf Exotic Fermions\footnote{to be published in {\it
Precision Tests of the Standard Model}, 1993, ed.\ Paul Langacker (World
Scientific).}}
\vskip 2truecm
\centerline{David London}
\centerline{Laboratoire de Physique Nucl\'eaire}
\centerline{Universit\'e de Montr\'eal}
\centerline{C.P. 6128, Montr\'eal, P. Q., Canada H3C 3J7}
\vskip 1truecm
\noindent
{\bf CONTENTS}
\vskip0.5truecm
\noindent
1. Introduction \dotfill 2

\noindent
2. Mixing Formalism \dotfill 4 \\
\null\qquad 2.1 Charged Fermions \dotfill 4 \\
\null\qquad 2.2 Neutrinos \dotfill 7

\noindent
3. Experimental Data \dotfill 10 \\
\null\qquad 3.1 $M_W$ \dotfill 11 \\
\null\qquad 3.2 Charged Currents \dotfill 11 \\
\null\qquad\qquad 3.2.1 Lepton Universality \dotfill 11 \\
\null\qquad\qquad 3.2.2 Quark-Lepton Universality \dotfill 12 \\
\null\qquad 3.3 Neutral Currents (Low Energy) \dotfill 13 \\
\null\qquad\qquad 3.3.1 Deep-Inelastic Neutrino Scattering \dotfill 13 \\
\null\qquad\qquad 3.3.2 Neutrino-Electron Scattering \dotfill 14 \\
\null\qquad\qquad 3.3.3 Atomic Parity Violation \dotfill 15 \\
\null\qquad 3.4 Neutral Currents ($Z$ Peak) \dotfill 15 \\
\null\qquad\qquad 3.4.1 $Z^0$ Decay Widths \dotfill 16 \\
\null\qquad\qquad 3.4.2 Leptonic Asymmetries \dotfill 16 \\
\null\qquad\qquad 3.4.3 Heavy Flavours \dotfill 16

\noindent
4. Constraints \dotfill 18

\noindent
5. Conclusions \dotfill 20

\vfill\eject

\section{Introduction}

As has often been said, the standard model, although extremely successful
in explaining virtually all known experimental data, cannot be the whole
story -- there are too many questions still left unanswered. By themselves,
exotic fermions do not solve any of these problems, but they often appear
in models which do address some of these remaining questions. In this
chapter I will discuss the constraints which can be put upon exotic
fermions, particularly as regards their mixings with the ordinary fermions.

In the standard model, all left-handed ($L$) fermions transform as doublets
under weak $\gweak$, while all right-handed ($R$) fermions are singlets:
\beq
{\nu_e\choose e^-}_L~~~{\nu_\mu\choose \mu^-}_L~~~
{\nu_\tau\choose \tau^-}_L~~~{u \choose d}_L~~~
{c \choose s}_L~~~{t \choose b}_L~,
\eeq
\beq
e^-_R~~~\mu^-_R~~~\tau^-_R~~~
{u_R\atop d_R}~~~{c_R\atop s_R}~~~{t_R\atop b_R}~.
\eeq
Many models which go beyond the standard model predict the existence of new
fermions which transform in a non-standard way under $\gweak$. In $E_6$
models, for example, in the 27-plet one finds, in addition to the ordinary
particles, vector singlet quarks and vector doublet leptons. Vector singlet
(doublet) fermions refer to particles whose $L$ and $R$ components both
transform as singlets (doublets) under $\gweak$. One also finds new
$\gweak$-singlet Weyl neutrinos in the 27-plet. Mirror fermions are another
type of exotic fermion, whose transformation properties under $\gweak$ are
opposite those of ordinary fermions, i.e.~left-handed singlets and
right-handed doublets. These appear, for instance, in grand unified
theories which include family unification \cite{mirror}.

The possibilities for new fermions are listed in Table 1. In the following
analysis, all particles whose $L$ and $R$ components obey the same
transformation properties as those in Eqs.~1 and 2 (i.e.~$L$-handed
doublets, $R$-handed singlets) will be called {\it ordinary}. These include
the standard fermions, as well as any new sequential (e.g.~fourth family)
fermions. Those particles whose $L$ and/or $R$ components transform
differently than those of ordinary fermions are called {\it exotic}. Note
that particles with noncanonical electric or colour charges are not
considered here. This restricts the `exotic' label to mirror fermions,
vector doublets and singlets, and Weyl neutrinos.

\begin{table}
\hfil
\vbox{\offinterlineskip
\halign{&\vrule#&
   \strut\quad#\hfil\quad\cr
\noalign{\hrule}
height2pt&\omit&\cr
&Sequential Fermions&\cr
&~~~~~~~${N\choose E^-}_L~~~E_R~,~~~{U\choose D}_L~~~{U_R\atop D_R}$&\cr
height2pt&\omit&\cr
\noalign{\hrule}
height2pt&\omit&\cr
&Non-Canonical $\gweak\times U(1)$ Assignments&\cr
&~~~~~a) Mirror Fermions&\cr
&~~~~~~~$E_L^-~~~{N\choose E^-}_R~,~~~{U_L\atop D_L}~~~{U\choose D}_R$&\cr
height2pt&\omit&\cr
&~~~~~b) Vector Doublets&\cr
&~~~~~~~${N\choose E^-}_L~~~{N\choose E^-}_R~,
      ~~~{U\choose D}_L~~~{U\choose D}_R$&\cr
height2pt&\omit&\cr
&~~~~~c) Vector Singlets&\cr
&~~~~~~~$E^-_L~~~E^-_R~,~~~U_L~~~U_R~~~D_L~~~D_R$&\cr
height2pt&\omit&\cr
&~~~~~d) Weyl Neutrinos&\cr
&~~~~~~~$N_L~~~N_R$&\cr
height2pt&\omit&\cr
\noalign{\hrule}}}
\caption{Possible $\gweak\times U(1)$ assignments for new fermions. Pairs
of particles enclosed in parentheses indicate $\gweak$-doublets; otherwise
they are $\gweak$-singlets. $N$ and $E$ refer to leptons of charge $0$ and
$-1$, respectively; $U$ and $D$ are quarks of charge $2/3$ and $-1/3$.}
\end{table}

There are two ways to look for signals of exotic fermions -- directly and
indirectly. The best limits on direct production of such particles come
from LEP \cite{PDG1992},
\beq
M_N,M_{E^-},M_U,M_D > 45~{\rm GeV},
\eeq
although the bound on $M_N$ depends on the type of exotic neutrino. For
example, the mass limit on exotic singlets can be considerably weaker. As
to indirect signals, one possibility is to look for loop-induced effects in
rare processes. This is a model-dependent enterprise, depending on the mass
of the exotic fermions, their couplings to ordinary gauge bosons, the
possible existence of other gauge bosons, etc., and will not be discussed
here.

The other indirect signal, which is the focus of this chapter, is to look
for signs of exotic fermions through their mixings with ordinary fermions.
These mixings can be analysed in a model independent way. In general,
mixing between ordinary and exotic fermions will induce flavour-changing
neutral currents (FCNC). The experimental absence of FCNC places extremely
stringent limits on fermion mixing. However, there are directions in
parameter space where it is possible to fine-tune away FCNC. Nevertheless,
even these regions can be constrained by looking at data involving charged
currents and flavour-conserving neutral currents. I will review here the
constraints which current experimental data place upon the mixings of
ordinary and exotic fermions. The material contained in this chapter comes
principally from work by Langacker and London (Ref.~\cite{LL}), and
Nardi, Roulet and Tommasini (Ref.~\cite{NRT}). Where there are holes in the
exposition, I refer the reader to these two articles for details.

This chapter is organized as follows. In section 2, I introduce the
formalism needed to describe the mixing between ordinary and exotic
fermions. For charged fermions, in order to avoid FCNC, it is necessary to
consider fine-tuned directions in parameter space in which each ordinary
charged fermion mixes with its own exotic fermion. In this way mixing is
parametrized by one angle per ordinary ($L$- or $R$-handed) charged
fermion. Neutrinos are more complicated, both due to the possibility of
Dirac and Majorana masses, and because there is no empirical evidence
requiring the absence of FCNC between neutrino species. However, due to the
fact that neutrinos are unobserved in experiment, it is possible to
parametrize mixing by one effective angle, plus one auxiliary parameter,
per neutrino species. In section 3, I review the experimental data which is
used to constrain mixing. Here I will include the theoretical expressions,
including mixing, which are to be fitted to the experimental results. The
fact that certain results are normalized to other data must be carefully
taken into account to ensure that {\it all} mixing effects are included.
The fits are given in section 4. There are enough constraints to limit all
mixings of $L$- and $R$-handed ordinary fermions. Two types of fits are
presented. In the first, only one particle at a time is allowed to mix.
This yields the most stringent limits on fermion mixing. In the second fit,
all particles can mix simultaneously, which weakens the constraints due to
the possibility of fine-tuned cancellations of the effects of different
particle mixings. I conclude in section 5.

\section{Mixing Formalism}

In this section, I present the formalism for describing the mixing of
ordinary and exotic fermions. As mentioned in the introduction, charged
fermions and neutrinos must be treated separately. The material in this
section is taken almost completely from Ref.~\cite{LL}.

\subsection{Charged Fermions}

Since electromagnetic gauge invariance is unbroken, fermions with different
charges cannot mix. Charged fermions can therefore be divided into three
categories -- those with $Q_{em}=2/3$ ($u$-type), $Q_{em}=-1/3$ ($d$-type)
and $Q_{em}=-1$ ($e$-type). For each of these types it is convenient to put
the $L$ and $R$ gauge eigenstates of both ordinary and exotic fermions into
a single vector,
\beq
\psi_{L(R)}^0=\pmatrix{\psi^0_O\cr\psi^0_E\cr}_{L(R)},
\eeq
in which the subscripts $O$ and $E$ stand for ordinary and exotic,
respectively. Here and below, the superscript $0$ indicates the
weak-interaction basis; the mass basis is denoted by the absence of
superscripts. In the above equation there are $n_L$ ($n_R$) ordinary
$L$-handed ($R$-handed) fields and $m_L$ ($m_R$) exotic $L$-handed
($R$-handed) fields.

The light ($l$) and heavy ($h$) mass eigenstates can be written similarly,
\beq
\psi_{L(R)}=\pmatrix{\psi_l\cr\psi_h\cr}_{L(R)}.
\eeq
The dimensionality of these vectors is as above, that is, there are $n_L$
($n_R$) light $L$ ($R$) states and $m_L$ ($m_R$) heavy $L$ ($R$) states. Of
course, the labels `light' and `heavy' should not be taken literally -- for
example, fourth generation particles (if they exist) are known to be heavy,
yet they are included among the `light' particles. This decomposition is
useful as a reminder that in general we expect the light states to consist
mainly of ordinary particles and the heavy states to be principally exotic.

The weak and mass eigenstates are related by a unitary transformation
\beq
\label{transone}
\psi_a^0=U_a\psi_a,
\eeq
in which $a=L,R$. It is useful to write the matrix $U$ in block form as
\beq
\label{trans}
U_a=\pmatrix{A_a & E_a \cr F_a & G_a \cr}.
\eeq
Since all our experimental data concerns only the light eigenstates, the
important elements of $U_a$ are the $n_a\times n_a$ submatrix $A_a$, which
relates the light mass states and the ordinary weak states, and $F_a$,
which is $m_a\times n_a$ and describes the overlap of the light eigenstates
with the exotic fermions. The unitarity of $U_a$ requires
\beq
\label{unitarity}
A_a^\dagger A_a + F_a^\dagger F_a = A_a A_a^\dagger + E_a E_a^\dagger = I,
\eeq
which shows that $A_a$ is not by itself unitary. However, since we expect
the light (heavy) particles to be mainly ordinary (exotic), we see that the
deviation of $A_a$ from unitarity is of second order in the mixing.

Let us now examine the effects of mixing on the neutral currents of the
light fermions. In the weak basis, the coupling of the $Z^0$ to charged
fermions can be written
\beq
{1\over 2}J_Z^\mu = {\overline\psi}^0_{OL} I_{3W} \gamma^\mu \psi^0_{OL}
+ {\overline\psi}^0_{ER} I_{3W} \gamma^\mu \psi^0_{ER}
- \sin^2\theta_W J^\mu_{em},
\eeq
in which $I_{3W}=+1/2$ for $u$-type fermions, and $I_{3W}=-1/2$ for $d$-
and $e$-type fermions. Using Eqs.~\ref{transone} and \ref{trans}, the weak
neutral current can now be expressed in terms of mass eigenstates. Keeping
only those terms which involve just the light states, this gives
\beq
\label{neutcurr}
{1\over 2}J_Z^\mu =
{\overline\psi}_{lL} \gamma^\mu I_{3W} A_L^\dagger A_L \psi_{lL}
+ {\overline\psi}_{lR} \gamma^\mu I_{3W} F_R^\dagger F_R \psi_{lR}
- {\overline\psi}_{l} \gamma^\mu Q_{em} \sin^2\theta_W \psi_{l}.
\eeq
The important point to recognize here is that, since neither $A_L$ nor
$F_R$ is unitary (Eq.~\ref{unitarity}), $A_L^\dagger A_L$ and $F_R^\dagger
F_R$ are not necessarily diagonal. In other words, FCNC will in general be
induced among the light particles.

It is useful to parametrize the FCNC between the light particles $i$ and
$j$ as
\beq
\label{lambda}
\lambda^L_{ij} = \left(A_L^\dagger A_L\right)_{ij}
= -\left(F_L^\dagger F_L\right)_{ij}~,~~~
\lambda^R_{ij} = \left(F_R^\dagger F_R\right)_{ij}~,~~~i\ne j.
\eeq
Note that these are of second order in light-heavy mixing. As can be seen
from Table 2, the constraints on the $\lambda^{L,R}_{ij},~i\ne j$ are quite
stringent, which strongly limits the mixing of ordinary and exotic
fermions. However, it is possible to evade these bounds by considering the
fine-tuned cases in which both $A_L^\dagger A_L$ and $F_R^\dagger F_R$ are
diagonal. These correspond to those directions in mixing parameter space in
which each ordinary fermion mixes with its own exotic fermion. For the rest
of this chapter, I will assume $\lambda^{L,R}_{ij}=0$ for $i\ne j$.

\begin{table}
\hfil
\vbox{\offinterlineskip
\halign{&\vrule#&
   \strut\quad#\hfil\quad\cr
\noalign{\hrule}
height2pt&\omit&&\omit&&\omit&\cr
& Quantity && Upper Limit && Source &\cr
height2pt&\omit&&\omit&&\omit&\cr
\noalign{\hrule}
height2pt&\omit&&\omit&&\omit&\cr
& $\vert\lambda_{\mu e}\vert$ && $1\times 10^{-6}$ &&
			$\mu\not\to 3e$ \cite{PDG1992} &\cr
& $\vert\lambda_{\mu\tau}\vert$, $\vert\lambda_{e\tau}\vert$
	&& $7\times 10^{-3}$ && $\tau\not\to 3\ell$ \cite{PDG1992} &\cr
& $\vert\lambda_{ds}\vert$ && $6\times 10^{-4}$ &&
			$\Delta m_{K_L K_S}$ \cite{PDG1992} &\cr
& \omit && $1\times 10^{-5}$ && $K_L\to\mu^+\mu^-$ \cite{PDG1992} &\cr
& $\vert\lambda_{cu}\vert$ && $1\times 10^{-3}$ &&
			$D^0$-${\overline{D^0}}$ mixing \cite{PDG1992} &\cr
& $\vert\lambda_{bd}\vert$, $\vert\lambda_{bs}\vert$
	&& $2\times 10^{-3}$ && $B\not\to\ell^+\ell^- X$ \cite{UA1-B} &\cr
height2pt&\omit&&\omit&&\omit&\cr
\noalign{\hrule}}}
\caption{Limits on the flavour changing neutral current parameters
$\lambda_{ij}$ (Eq.~\protect\ref{lambda}). The bounds on leptonic FCNC are
taken or adapted from Ref.~\protect\cite{Nardi}, while the limits on
hadronic FCNC are taken, updated or adapted from
Ref.~\protect\cite{Silverman}. There is no bound on
$\vert\lambda_{bd}\vert$ from $B_d^0$-${\overline{B_d^0}}$ mixing because
this mixing can in principle be explained by a nonzero $\lambda_{bd}$
\protect\cite{Silverman}.}
\end{table}

With this (strong) assumption, using Eq.~\ref{unitarity} one can write
\beq
\left(A_a^\dagger A_a\right)_{ij}=\left(c^i_a\right)^2 \delta_{ij}~,~~~~~
\left(F_a^\dagger F_a\right)_{ij}=\left(s^i_a\right)^2 \delta_{ij}~,~~~~~
{}~~~~~~a=L,R,
\eeq
in which $\left(s^i_a\right)^2 \equiv 1 - \left(c^i_a\right)^2 \equiv
\sin^2\theta_a^i$, where $\theta_{L(R)}^i$ is the mixing angle of the
$i^{th}$ $L$-handed ($R$-handed) ordinary fermion and its exotic partner.
With this notation, the neutral current in Eq.~\ref{neutcurr} becomes
\beq
\label{neutral}
{1\over 2}J_Z^\mu = \sum_i\left[
{\overline\psi}_{iL} \gamma^\mu {\tilde\epsilon}_L(i) \psi_{iL}
+ {\overline\psi}_{iR} \gamma^\mu {\tilde\epsilon}_R(i) \psi_{iR} \right],
\eeq
where the sum is over the light particles and
\begin{eqnarray}
\label{epslr}
{\tilde\epsilon}_L(i) & = & I_{3W}^i \clisq - Q^i_{em} \sin^2\theta_W~,
\nonumber \\
{\tilde\epsilon}_R(i) & = & I_{3W}^i \srisq - Q^i_{em} \sin^2\theta_W~.
\end{eqnarray}
{}From Eqs.~\ref{neutral} and \ref{epslr}, the effects of mixing are clear.
First, the mixing of ordinary $L$ doublets with exotic $L$ singlets results
in a nonuniversal reduction ($\left(c^i_L\right)^2$) of the isospin
current. Second, mixing in the $R$-handed sector induces a $R$-handed
current ($\left(s^i_R\right)^2$). The electromagnetic current is unchanged,
reflecting the simple fact that only particles of the same charge can mix.
In the presence of mixing, the vector and axial couplings for fermion $i$
are
\begin{eqnarray}
\label{vecax}
v_i & \equiv & {\tilde\epsilon}_L(i) + {\tilde\epsilon}_R(i) =
I_{3W}^i \left[ \clisq + \srisq \right] - 2 Q^i_{em} \sin^2\theta_W~,
\nonumber \\
a_i & \equiv & {\tilde\epsilon}_L(i) - {\tilde\epsilon}_R(i) =
I_{3W}^i \left[ \clisq - \srisq \right].
\end{eqnarray}

The hadronic charged current involving the light quarks is
\beq
\label{charged}
{1\over 2}J_W^{\mu^\dagger} =
{\overline\psi}_{uL} \gamma^\mu V_L \psi_{dL}
+ {\overline\psi}_{uR} \gamma^\mu V_R \psi_{dR},
\eeq
in which $\psi_{uL}$ and $\psi_{dL}$ are column vectors of the light $L$
$u$-type and $d$-type quarks, respectively. Recall that `light' $L$-handed
particles include possible extra sequential or vector doublet quarks. Thus,
the first 3 components of $\psi_{uL}$ and $\psi_{dL}$ are the standard
quarks, while the remaining $n_l-3$ quarks are nonstandard. The column
vectors $\psi_{uR}$ and $\psi_{dR}$ are defined completely analogously. In
Eq.~\ref{charged}, $V_L=A_L^{u^\dagger}A_L^d$ is the generalized
Cabibbo-Kobayashi-Maskawa (CKM) matrix. The point to observe here, however,
is that $V_L$ is non-unitary in the presence of mixing between the ordinary
and exotic fermions. It can be decomposed as
\beq
\label{vlckm}
V_{Lij} = c_L^{u_i} c_L^{d_j} {\widehat V}_{Lij},
\eeq
where ${\widehat V}_L$ is the true (unitary) CKM matrix. Here and below I
use the term `true' to refer to a quantity in the absence of mixing, and I
denote this by a symbol with a caret. `Apparent' quantities, which are
represented by symbols with no caret, are those which are actually
measured. In Eq.~\ref{vlckm}, we see that apparent CKM matrix elements are
reduced from their true values by the nonuniversal factor $c_L^{u_i}
c_L^{d_j}$. If $\psi_{uL}$ and/or $\psi_{dL}$ contain nonstandard `light'
quarks, this will manifest itself through the apparent nonunitarity of
${\widehat V}_L$. The second term in Eq.~\ref{charged} is a $R$-handed
charged current, induced when both $R$-handed $u_i$ and $d_j$ quarks mix
with exotic $\gweak$-doublets. Like $V_L$, the apparent $R$-handed CKM
matrix $V_R$ is non-unitary, but can be written
\beq
\label{vrckm}
V_{Rij} = s_R^{u_i} s_R^{d_j} {\widehat V}_{Rij},
\eeq
where ${\widehat V}_R$ is unitary.

\subsection{Neutrinos}

As mentioned in the introduction, neutrinos must be treated separately for
several reasons. First of all, there are three types of $L$-handed neutrino
weak eigenstates:
\beq
\label{nutypes}
\pmatrix{n_{OL}^0 \cr e_L^{0^-}\cr}~,~~~~~
\pmatrix{e_L^{0^+}\cr n_{EL}^0 \cr}~,~~~~~n_{SL}^0~.
\eeq
Here, the $n_{OL}^0$ are ordinary $\gweak$-doublets with $I_{3W}=1/2$, the
$n_{EL}^0$ are exotic $\gweak$-doublets with $I_{3W}=-1/2$, and the
$n_{SL}^0$ are exotic $\gweak$-singlets. Note that the $n_{EL}^0$ are
usually referred to as antineutrinos. However, Majorana masses are possible
for neutrinos, in which case there is no real distinction between particle
and antiparticle. This is the second difference between neutrinos and
charged particles. In the general Majorana case all three types of $\nu$
can mix. Finally, there are no experimental constraints on FCNC involving
neutrinos. Despite these differences, mixing between ordinary and exotic
neutrinos can be analyzed using a formalism similar to that introduced in
Sec.~2.1.

Since in the presence of Majorana masses one does not distinguish between
particle and antiparticle, in dealing with neutrinos it is convenient to
denote all $L$ states as $n_L$ and all $R$ states as $n_R^c$. These are
related by $n_R^c=C({\overline n}_L^T)$, where $C$ is the charge
conjugation matrix. Thus, in analogy to the charged fermion case, all
$L$-handed weak eigenstate neutrinos are put together into a vector
\beq
n_L^0=\pmatrix{n_{OL}^0 \cr n_{EL}^0 \cr n_{SL}^0}.
\eeq
As above, the neutrino mass eigenstates are divided into two classes,
`light' (i.e.~essentially massless) and `heavy':
\beq
n_L=\pmatrix{n_{lL}\cr n_{hL}\cr}.
\eeq
The weak and mass bases are related by a unitary transformation
$n_L^0 = U_L n_L$, in which $U$ can be decomposed as
\beq
\label{nutrans}
U_L=\pmatrix{A & E \cr F & G \cr H & J \cr }_L~.
\eeq
Similarly, $n_R^{0^c} = U_R n_R^c$, with $U_R=U_L^*$. In Eq.~\ref{nutrans},
the matrices $A_L$, $F_L$ and $H_L$ describe the overlap of the massless
neutrinos with ordinary doublets ($n_{OL}^0$), exotic doublets
($n_{EL}^0$), and exotic singlets ($n_{SL}^0$), respectively. The LEP data
has constrained the number of light $\gweak$-doublets to be 3. Thus, exotic
doublet $\nu$'s must have a mass greater than $M_Z/2$. This implies that
the components of $F_L$ are small. As to $H_L$, I will assume that the
light neutrinos are mainly $n_{OL}^0$. If they are massless or have
Majorana masses, then there are no light singlets, and all components of
$H_L$ are small. If the $n_{lL}$ have small Dirac masses, then it is
necessary to include 3 light singlets in the spectrum. In this case, the
components of $H_L$ corresponding to these singlets may be large, but the
remaining components must be small. As far as the formalism is concerned,
there is little difference between these two possibilities.

Dropping the subscript $l$, the weak neutral current for the light neutrino
states can now be written
\beq
\label{nuneutral}
{1\over 2}J_Z^\mu = {1\over 2}{\overline n}_L \gamma^\mu
\left( A_L^{\nu^\dagger} A_L^\nu - F_L^{\nu^\dagger} F_L^\nu \right) n_L~.
\eeq
The $A_L^{\nu^\dagger} A_L^\nu$ and $F_L^{\nu^\dagger} F_L^\nu$ terms come
from the neutral currents of the $n_{OL}^0$ and $n_{EL}^0$, respectively.
As in Sec.~2.2, neither $A_L$ nor $F_L$ is unitary. On the other hand,
unlike the charged fermion case, there is no experimental evidence to
suggest that $A_L^{\nu^\dagger} A_L^\nu$ and $F_L^{\nu^\dagger} F_L^\nu$
are diagonal. However, as we will see, essentially the same effect is
produced when one sums over the unobserved final state $\nu$'s in weak
processes.

The leptonic charged current is
\begin{eqnarray}
\label{nucharged}
{1\over 2}J_W^{\mu^\dagger} & = &
{\overline n}_L \gamma^\mu A_L^{\nu^\dagger} c_L^e e_L
+ {\overline n}_R^c \gamma^\mu F_R^{\nu^\dagger} s_R^e e_R \nonumber \\
& = & \sum_{ia} \left[ {\overline n}_{iL} \gamma^\mu
\left(A_L^{\nu^\dagger}\right)_{ia} c_L^{e_a} e_{aL}
+ {\overline n}_{iR}^c \gamma^\mu
\left(F_R^{\nu^\dagger}\right)_{ia} s_R^{e_a} e_{aR} \right].
\end{eqnarray}
Note that since $F_R^\nu = F_L^{\nu^*}$, the second term in
Eq.~\ref{nucharged}, which is the induced right-handed current, is of
second order in light-heavy mixing. This term is produced when both the
light neutrino and charged lepton mix with a member of an exotic doublet.
The left-handed charged current is reduced in strength by the factor
$\left(A_L^{\nu^\dagger}\right)_{ia} c_L^{e_a}$ due to ordinary-exotic
mixing.

We can now see the effect of summing over the final state $\nu$'s in a weak
process. In the presence of mixing, the rate for the charged current
transition $e_a\to n_i$ relative to its value ($\Gamma_0$) in the absence
of mixing is
\beq
{1\over \Gamma_0} \Gamma(e_a\to n_i) = \left(c_L^{e_a}\right)^2
\left(A_L^\nu\right)_{ai} \left(A_L^{\nu^\dagger}\right)_{ia} +
\left(s_R^{e_a}\right)^2
\left(F_R^\nu\right)_{ai} \left(F_R^{\nu^\dagger}\right)_{ia}.
\eeq
However, since the final $\nu$'s are unobserved, we must sum over them. The
effect of this is to reduce the many parameters describing neutrino mixing
to a single mixing angle per neutrino flavour:
\beq
\label{nusumrate}
{1\over \Gamma_0} \sum_i \Gamma(e_a\to n_i) =
\left(c_L^{e_a}\right)^2 \left(c_L^{\nu_a}\right)^2 +
\left(s_R^{e_a}\right)^2 \left(s_R^{\nu_a}\right)^2~,
\eeq
where the effective neutrino mixing angles $\left(c_L^{\nu_a}\right)^2 =
\left(A_L^\nu A_L^{\nu^\dagger}\right)_{aa}$ and
$\left(s_R^{\nu_a}\right)^2 = \left(F_R^\nu F_R^{\nu^\dagger}\right)_{aa}$
have been introduced. The second term in Eq.~\ref{nusumrate}, which comes
from the induced right-handed charged current, is of $O(s^4)$. From now on
we will be working to second order in light-heavy mixing, so that this term
can be dropped.

The final state neutrino produced in Eq.~\ref{nusumrate} is
\beq
\label{nustate}
\ket{n_{aL}} \equiv { \sum_i \left(A_L^{\nu^\dagger}\right)_{ia}
\ket{n_{iL}} \over c_L^{\nu_a} },
\eeq
so that the cross section for scattering into the ``right'' charged lepton
($e_{aL}$) is
\beq
{1\over \sigma_0} \sigma(n_{aL}\to e_{aL}) =
\left(c_L^{e_a}\right)^2 \left(c_L^{\nu_a}\right)^2~.
\eeq
(There is also the possibility of scattering into the ``wrong'' lepton
(Ref.~\cite{nonorthogonal}), but this will not be discussed here.) One can
also calculate the neutral current cross section for the rescattering of
the neutrino in Eq.~\ref{nustate}. Summing again over the final unobserved
neutrinos, this can be found from Eqs.~\ref{nuneutral} and \ref{nustate} to
give
\begin{eqnarray}
\label{nurescatter}
{1\over \sigma_0} \sum_i \sigma(n_{aL}\to n_{iL}) & = &
{ 1 \over \left(c_L^{\nu_a}\right)^2} \left[ A_L^\nu
\left( A_L^{\nu^\dagger} A_L^\nu - F_L^{\nu^\dagger} F_L^\nu \right)^2
A_L^{\nu^\dagger} \right]_{aa} \nonumber \\
& = & 1 - {2 \over \left(c_L^{\nu_a}\right)^2} \left[ A_L^\nu
\left( 2 F_L^{\nu^\dagger} F_L^\nu + H_L^{\nu^\dagger} H_L^\nu \right)
A_L^{\nu^\dagger} \right]_{aa} + O(s^4),
\end{eqnarray}
where, in the second line, I have used the unitarity of $U_L$
(Eq.~\ref{nutrans}) and the fact that the components of $F_L$ and $H_L$ are
all of $O(s)$\footnote{I have assumed that the light $\nu$'s are either
massless or Majorana; the case of light Dirac $\nu$'s does not change the
formalism significantly.}. Thus it is evident that this cross section
depends not only on the mixing angle, but also on the type of neutrino(s)
with which the ordinary neutrino mixes. Eq.~\ref{nurescatter} simplifies
even further when one realizes that, for $a \epsilon 1,2,...,p$ ($p$ is the
number of light $\nu$'s), $A_L^\nu$ differs from the identity by terms of
$O(s)$. In this case we obtain
\beq
{1\over \sigma_0} \sum_i \sigma(n_{aL}\to n_{iL}) =
1 - \Lambda_a \left(s_L^{\nu_a}\right)^2,
\eeq
where the parameter $\Lambda_a$ is defined to be $\Lambda_a = 4\lambda_F^a
+ 2\lambda_H^a$, with $(F_L^{\nu^\dagger} F_L^\nu)_{aa} \equiv
\lambda_F^a\left(s_L^{\nu_a}\right)^2$ and $(H_L^{\nu^\dagger}
H_L^\nu)_{aa} \equiv \lambda_H^a\left(s_L^{\nu_a}\right)^2$. The
$\lambda$'s are constrained to lie between 0 and 1, so that $\Lambda_a$
takes values between 0 and 4, depending on the mixing involved.

Finally, using Eq.~\ref{nuneutral} and the same approximations as above, it
is possible to calculate the rate for the decay of the $Z^0$ into
undetected neutrinos. Assuming the existence of 3 light neutrinos in the
absence of mixing, and normalizing to the decay rate of the $Z^0$ into one
neutrino, this gives
\beq
{1 \over \Gamma_0^{1\nu}} \Gamma(Z^0\to invisible) =
3 - \sum_a \Lambda_a \left(s_L^{\nu_a}\right)^2.
\eeq

Having presented the formalism for the mixing of ordinary and exotic
fermions, I will now turn to the experimental data which is used to
constrain such mixings.

\section{Experimental Data}

In this section, I will present the experimental data which are used to
constrain the mixing of ordinary and exotic fermions. I must emphasize at
the outset that these results, taken from Ref.~\cite{NRT}, are somewhat
outdated, since the analysis was done in the summer of 1991. However,
except for the $\nu_L^\tau$, the constraints obtained here would not be
much improved if present data were used. I will comment further on this in
Section 4.

In using the data to constrain fermion mixing, it must be remembered that
mixing can cause a discrepancy between the experimental result and the
theoretical expression in two ways. Not only can mixing directly affect the
process being examined, but it can also appear indirectly. This can happen,
for example, when the extraction of a particular result requires
normalization to another piece of experimental data. Thus, in putting
constraints on mixing, one must be very careful to include {\it all} mixing
effects.

Most of the experimental results are precise enough that it is necessary to
include radiative corrections in order that there be agreement with the
standard model. In the present analysis, radiative corrections will be
included, but only those due to ordinary particles without mixing. (The
inclusion of mixing is a second order effect.) Radiative corrections
involving exotic fermions are typically much smaller, although it must be
acknowledged that in the case of exotic nondegenerate $\gweak$ doublets,
the corrections could be large \cite{veltman}.

In order to calculate radiative corrections, it is necessary to choose a
set of input parameters. These are typically taken to be the
electromagnetic coupling $\alpha$, measured at $q^2=0$, the Fermi constant
$G_\mu$, and the $Z$-mass $M_Z$, fixed to be $M_Z=91.175$ GeV \cite{MZ}.
The values of $\alpha$ and $M_Z$ as extracted from experiment are not
affected by mixing. On the other hand, since $G_\mu$ is obtained directly
from $\mu$-decay, there is an effect due to mixing. The measured value of
$G_\mu=1.16637(2)\times 10^{-5}~{\rm GeV}^{-2}$ is related to its true
value ${\widehat G}_\mu$ by
\beq
\label{gfermi}
G_\mu = {\widehat G}_\mu \cle\clnue\clmu\clnumu
\eeq
due to the possible mixing of the leptons with exotic fermions. Since many
experimental results are normalized to $\mu$-decay, indirect effects of
mixing can appear in this way. Finally, it is necessary to include
$t$-quark mass and the Higgs mass in the radiative corrections. These are
fixed to be $m_t=120$ GeV and $m_H=100$ GeV, respectively.

\subsection{$M_W$}

Including radiative corrections, the theoretical expression for $M_W$ as a
function of $\alpha$, $M_Z$ and $G_\mu$ is given by
\cite{MWradcorr},\cite{radcorr}
\beq
\label{wmass}
M_W^2 = {\rho M_Z^2 \over 2} \left[ 1 + \sqrt { 1 -
{ G_\mu \over {\widehat G}_\mu } { 4 {\cal A} \over \rho M_Z^2 }
\left( { 1 \over 1 - \Delta\alpha} + \Delta r^{rem} \right) } \, \right] ,
\eeq
where ${\cal A}=\pi\alpha/\sqrt{2} G_\mu$. Here, $\rho\simeq 1 + 3 G_\mu
m_t^2/8\sqrt{2}\pi^2$ contains the leading $t$-quark effects
\cite{veltman}, $1/(1-\Delta\alpha)$ renormalizes the QED coupling to the
$M_Z$ scale, including the large logs, and $\Delta r^{rem}$ includes all
remaining small corrections. The sole (indirect) dependence of $M_W$ on
fermion mixings is found in the ratio $G_\mu / {\widehat G}_\mu$
(Eq.~\ref{gfermi}).

The average value of $M_W$ as measured by CDF and UA2 is \cite{MW}
\beq
M_W=80.13\pm0.31~{\rm GeV},
\eeq
where the LEP result for $M_Z$ has been used to convert the UA2 measurement
of $M_W/M_Z$ into a value for $M_W$.

\subsection{Charged Currents}

There are a number of experiments involving charged currents which can be
used to constrain fermion mixing. In the interest of brevity, I will
present only those experimental results which are most important for
bounding the mixing.

\subsubsection{Lepton Universality}

In the standard model, the coupling of the $W$ to each of the lepton
doublets $(\nu_e~e^-)_L$, $(\nu_\mu~\mu^-)_L$ and $(\nu_\tau~\tau^-)_L$ is
universal, that is, $g_e=g_\mu=g_\tau$. In the presence of mixing, this
equality can be altered:
\begin{eqnarray}
\label{luniversality}
\left({g_i\over g_e}\right)^2 & = & { \cllisq\clnuisq + \srlisq\srnuisq
\over \clesq\clnuesq + \sresq\srnuesq } \nonumber \\
& \simeq & 1 + \slesq + \slnuesq - \sllisq - \slnuisq~,~~~~~~~i=\mu,\tau,
\end{eqnarray}
where only terms of $O(s^2)$ have been kept in the second line.

These ratios have been measured in several experiments. The most precise
are
\begin{itemize}
\item Pion and Kaon decay:
\beq
{\Gamma(\pi\to\mu\nu) \over \Gamma(\pi\to e\nu) }~,~~~~~
{\Gamma(K\to\mu\nu) \over \Gamma(K\to e\nu) }~,
\eeq
\item Tau and Muon decay:
\beq
{\Gamma(\tau\to\mu{\bar\nu}\nu) \over \Gamma(\tau\to e{\bar\nu}\nu) }~,~~~~~
{\Gamma(\tau\to\mu{\bar\nu}\nu) \over \Gamma(\mu\to e{\bar\nu}\nu) }~.
\eeq
\end{itemize}
The experimental data are shown in Table 3. These experiments constrain the
left-handed mixing angles of the leptons. However, muon and tau decay have
been measured accurately enough to put limits on the right-handed mixing
angles of these leptons. The observables relevant to right-handed leptonic
currents are all of $O(s^4)$. I will not discuss these here, but rather
refer the reader to Refs.~\cite{LL} and \cite{mudecay} for details.

\begin{table}
\hfil
\vbox{\offinterlineskip
\halign{&\vrule#&
   \strut\quad#\hfil\quad\cr
\noalign{\hrule}
height2pt&\omit&&\omit&&\omit&\cr
& Quantity && Measured Value && Source &\cr
height2pt&\omit&&\omit&&\omit&\cr
\noalign{\hrule}
height2pt&\omit&&\omit&&\omit&\cr
& $(g_\mu/g_e)^2$ && $1.014\pm 0.011$ && $\pi\to\ell\nu$ \cite{PDG1990} &\cr
& $(g_\mu/g_e)^2$ && $1.013\pm 0.046$ && $K\to\ell\nu$ \cite{PDG1990} &\cr
& $(g_\mu/g_e)^2$ && $1.016\pm 0.026^\dagger$ &&
$\Gamma(\tau\to\mu{\bar\nu}\nu)/\Gamma(\tau\to e{\bar\nu}\nu)$
		\cite{PDG1990},\cite{taudecay} &\cr
& $(g_\tau/g_e)^2$ && $0.952\pm 0.031^\dagger$ &&
$\Gamma(\tau\to\mu{\bar\nu}\nu)/\Gamma(\mu\to e{\bar\nu}\nu)$
		\cite{PDG1990},\cite{taudecay} &\cr
height2pt&\omit&&\omit&&\omit&\cr
\noalign{\hrule}}}
\caption{Experimental constraints on lepton universality
(Eq.~\protect\ref{luniversality}). There is a correlation between the data
marked with a $^\dagger$, which has been taken into account in the fits
\protect\cite{NRT}.}
\end{table}

\subsubsection{Quark-Lepton Universality}

In order to test quark-lepton universality, one typically compares the rate
for the decay of a hadron with that of muon decay. In the standard model,
in the absence of mixing, these should be equal, up to factors of CKM
matrix elements. However, these matrix elements obey another constraint,
namely that of the unitarity of the CKM matrix. In this sense, a test of
quark-lepton universality is equivalent to a test of CKM matrix unitarity.

$V_{ud}$ is measured by comparing the rates for $\beta$-decay (vector
current only) and $\mu$-decay. $V_{us}$ is obtained similarly, except that
$K_{e3}$ and hyperon decay are used. In the presence of mixing, the true
values of the CKM matrix elements differ from the measured values by
\cite{LL}
\beq
\label{vui}
V_{ui} = {\clu\cli{\widehat V}_{Lui} + \sru\sri{\widehat V}_{Rui} \over
\clmu\clnumu}~~~~~~~i=d,s.
\eeq
$V_{ub}$ is also related to its true value in this way, but in any case its
size is too small to be of interest for this analysis.

Using the fact that $\sum_{i=1}^n\vert{\widehat V}_{Lui}\vert^2=1$,
expanding Eq.~\ref{vui} to $O(s^2)$, and defining
\beq
\kappa_{ij} = s_R^{u_i} s_R^{d_j} {{\widehat V}_{Rij}\over{\widehat V}_{Lij}},
\eeq
one obtains
\begin{eqnarray}
\sum_{i=1}^3\vert V_{ui} \vert^2 = 1 + \slmusq + \slnumusq - \slusq -
\sum_{i=4}^n \vert{\widehat V}_{Lui}\vert^2 & + & \vert V_{ud}\vert^2
\left(2{\rm Re}(\kappa_{ud}) - \sldsq \right) \nonumber\\
& + & \vert V_{us}\vert^2 \left(2{\rm Re}(\kappa_{us}) - \slssq \right).
\end{eqnarray}
The experimental value for this quantity is \cite{PDG1990}
\beq
\sum_{i=1}^3\vert V_{ui} \vert^2 = 0.9981 \pm 0.0021.
\eeq

For those CKM matrix elements involving the $c$-quark, the analysis is
similar to the above, except that the mixing of the first-generation
particles can be neglected since such mixings are constrained considerably
better from other processes than from the relatively imprecise measurements
of $V_{cd}$ and $V_{cs}$. Thus we have
\begin{eqnarray}
V_{cd} & = & \clc{\widehat V}_{Lcd}~, \nonumber \\
V_{cs} & = & \clc\cls{\widehat V}_{Lcs} + \src\srs{\widehat V}_{Rcs} ~,
\end{eqnarray}
and
\beq
\sum_{i=1}^3\vert V_{ci} \vert^2 = 1 - \slcsq -
\sum_{i=4}^n \vert{\widehat V}_{Lci}\vert^2 + \vert V_{cs}\vert^2
\left(2{\rm Re}(\kappa_{cs}) - \slssq \right),
\eeq
where $\vert V_{cb}\vert^2$ has been neglected. The experimental value is
\cite{PDG1990}
\beq
\sum_{i=1}^3\vert V_{ci} \vert^2 = 1.08 \pm 0.37.
\eeq

The hadronic right-handed currents $\kappa_{ud}$ and $\kappa_{us}$ are
constrained through the unitarity of the CKM matrix. There are additional,
very stringent constraints coming from the predictions of PCAC for
nonleptonic $K_{\pi 3}$ amplitudes relative to $K_{\pi 2}$ amplitudes
\cite{RHC}. Interpreting these limits as 1$\sigma$ errors \cite{LL}, one
has
\beq
\kappa_{ud},\kappa_{us}=0\pm 0.0037.
\eeq
There are also additional (weak) constraints on $\kappa_{cd}$ and
$\kappa_{cs}$, but they will not be discussed here (see Refs.~\cite{LL} and
\cite{NRT}).

\subsection{Neutral Currents (Low Energy)}

At low energy, neutral current interactions can be parametrized through
effective lagrangians in which the $Z^0$ has been integrated out. In this
subsection, I will discuss three types of scattering processes -- $\nu q$,
$\nu e$ and $eq$. In all three cases, radiative corrections are important
\cite{NCradcorr}. For simplicity, these corrections are not shown
explicitly, but are included in the fits.

\subsubsection{Deep-Inelastic Neutrino Scattering}

The effective lagrangian describing the scattering of neutrinos from quarks
can be written as
\beq
-{\cal L}^{\nu q} = {4 G_F \over \sqrt{2}} {\bar\nu}_L \gamma^\mu \nu_L
\sum_{i=u,d,...}
\left[ \epsilon_L(i) {\overline q}_L^i \gamma_\mu q_L^i
+ \epsilon_R(i) {\overline q}_R^i \gamma_\mu q_R^i \right].
\eeq
In order to extract the values of $\epsilon_L(i)$ and $\epsilon_R(i)$, the
neutral current process are normalized to the corresponding charged current
processes, that is, the ratios
\beq
R_\nu = {\sigma(\nu N \to \nu X)\over \sigma(\nu N \to \mu^- X) }~,~~~~~
R_{\bar\nu} = {\sigma({\bar\nu} N \to {\bar\nu} X)\over
\sigma({\bar\nu} N \to \mu^+ X) }
\eeq
are used. Thus, mixing effects enter both in the numerator and in the
denominator. Taking all effects into account, the values of $\epsilon_L(i)$
and $\epsilon_R(i)$ obtained from deep-inelastic neutrino scattering are
\beq
\epsilon_{L,R}(i) = F_1(s^2,\kappa) {\tilde\epsilon}_{L,R}(i),
\eeq
where the ${\tilde\epsilon}_{L,R}(i)$ are defined in Eq.~\ref{epslr}, and
\cite{LL}
\beq
\label{fone}
F_1(s^2,\kappa) = {1 - {1\over 2}\Lambda_\mu\slnumusq \over
1 - \slmusq - \slnumusq - {\rm Re}(\kappa_{ud}) }
\eeq
incorporates the mixing effects in the neutrinos as well as in the
normalization. The experimental values of
$g_a^2\equiv\epsilon_a(u)^2+\epsilon_a(d)^2$ and
$\theta_a\equiv\tan^{-1}\left[\epsilon_a(u)/\epsilon_a(d)\right],~a=L,R$
are given in Table 4.

\begin{table}
\hfil
\vbox{\offinterlineskip
\halign{&\vrule#&
   \strut\quad#\hfil\quad\cr
\noalign{\hrule}
height2pt&\omit&&\omit&&\omit&\cr
& Quantity && Experimental Value && Source &\cr
height2pt&\omit&&\omit&&\omit&\cr
\noalign{\hrule}
height2pt&\omit&&\omit&&\omit&\cr
& $g_L^2$ && $0.2977\pm 0.0042$ && \omit &\cr
& $g_R^2$ && $0.0317\pm 0.0034$ && \omit &\cr
& $\theta_L$ && $2.50\pm 0.03$ && \omit &\cr
& $\theta_R$ && $4.59^{+0.44}_{-0.27}$ && Deep inelastic \cite{PDG1990} &\cr
height2pt&\omit&&\omit&&\omit&\cr
\noalign{\hrule}
height2pt&\omit&&\omit&&\omit&\cr
& $g_V^e$ && $-0.10\pm 0.05$ && Low-energy $\nu_\mu e$: &\cr
& $g_A^e$ && $-0.50\pm 0.04$ && BNL \cite{BNL} &\cr
height2pt&\omit&&\omit&&\omit&\cr
\noalign{\hrule}
height2pt&\omit&&\omit&&\omit&\cr
& $g_V^e$ && $-0.06\pm 0.07$ && High-energy $\nu_\mu e$: &\cr
& $g_A^e$ && $-0.57\pm 0.07$ && CHARM I \cite{CHARMI} &\cr
height2pt&\omit&&\omit&&\omit&\cr
\noalign{\hrule}
height2pt&\omit&&\omit&&\omit&\cr
& $g_V^e/g_A^e$ && $0.047\pm 0.046$ && CHARM II \cite{CHARMII} &\cr
height2pt&\omit&&\omit&&\omit&\cr
\noalign{\hrule}
height2pt&\omit&&\omit&&\omit&\cr
& $C_{1u}$ && $-0.249\pm 0.066^\dagger$ && \omit &\cr
& $C_{1d}$ && $0.391\pm 0.059^\dagger$ && Atomic parity \cite{atomic} &\cr
height2pt&\omit&&\omit&&\omit&\cr
\noalign{\hrule}
height2pt&\omit&&\omit&&\omit&\cr
& $C_{2u}-{1\over 2}C_{2d}$ && $0.21\pm 0.37$ && SLAC $e$-$D$ \cite{SLAC} &\cr
height2pt&\omit&&\omit&&\omit&\cr
\noalign{\hrule}}}
\caption{Low-energy neutral current data. There are non-negligible
correlations between the measurements marked with a $^\dagger$. These have
been taken into account in the fits \protect\cite{NRT}.}
\end{table}

\subsubsection{Neutrino-Electron Scattering}

The neutral current interaction of $\nu_\mu$ and $e$ can be described
by
\beq
-{\cal L}^{\nu_\mu e} = {2 G_F \over \sqrt{2}} {\bar\nu}_L \gamma^\mu \nu_L
{\bar e} \gamma_\mu \left(g_V^e-g_A^e \gamma_5\right)e.
\eeq
As in deep-inelastic neutrino scattering, the vector- and axial-couplings
of the electron are obtained by normalizing the neutral current process (in
this case $\nu_\mu$-$e$ scattering) to a charged current process
($\nu_\mu$-$hadron$ scattering). Again, mixing effects appear in both
places. The low-energy experiments from BNL normalize to the quasielastic
process $\nu_\mu n \to \mu^- p$, leading to
\begin{eqnarray}
\label{gvga}
g_V^e & = & F_2(s^2) v_e =
F_2(s^2)\left[-{1\over 2}\clesq -{1\over 2}\sresq + 2
\sin^2\theta_W \right], \nonumber \\
g_A^e & = & F_2(s^2) a_e =
F_2(s^2)\left[-{1\over 2}\clesq +{1\over 2}\sresq\right],
\end{eqnarray}
where \cite{LL}
\beq
F_2(s^2) = {1 - {1\over 2}\Lambda_\mu\slnumusq\over 1 - \slmusq - \slnumusq}.
\eeq
The high-energy experiments at CERN and Fermilab normalize to $\nu N \to
\mu^- X$ as in deep-inelastic scattering, so that in this case $g_V^e$ and
$g_A^e$ are as in Eq.~\ref{gvga}, but with $F_2(s^2)$ replaced by
$F_1(s^2,\kappa)$ of Eq.~\ref{fone}. The experimental values of $g_V^e$ and
$g_A^e$ are shown in Table 4. Note that the CHARM II collaboration has
recently measured the ratio $g_V^e/g_A^e$, in which the dependence on
$F_1(s^2,\kappa)$ cancels.

\subsubsection{Atomic Parity Violation}

Atomic parity violation arises through the interference of the
electromagnetic and weak interactions. The parity violating couplings
$C_{1i}$ and $C_{2i}$ are defined by
\beq
-{\cal L}^{e q} = {G_F \over \sqrt{2}} \sum_i \left[ C_{1i} \,
{\bar e}_L \gamma_\mu \gamma_5 e \, {\overline q}^i \gamma^\mu q^i + C_{2i} \,
{\bar e}_L \gamma_\mu e \, {\overline q}^i \gamma^\mu \gamma_5 q^i \right].
\eeq
Including mixing, these couplings are given by
\beq
C_{1i}=2 \left({{\widehat G}_\mu\over G_\mu}\right) a_e v_i~,~~~~~
C_{2i}=2 \left({{\widehat G}_\mu\over G_\mu}\right) v_e a_i~,
\eeq
where the vector and axial couplings have been defined in Eq.~\ref{vecax}
and ${\widehat G}_\mu/G_\mu$ in Eq.~\ref{gfermi}. $C_{1u}$ and $C_{1d}$ are
measured in parity violating transitions in cesium; the combination
$C_{2u}-{1\over 2}C_{2d}$ has been determined in polarized $e$-$D$
scattering at SLAC. All the experimental values are given in Table 4.

\subsection{Neutral Currents ($Z$ Peak)}

The very accurate measurements at LEP put strong constraints on the mixing
of ordinary and exotic fermions, particularly as regards the $\tau$-lepton
and heavy quarks. Here I will present the experimental data on the decay
widths of the $Z^0$ as well as the forward-backward asymmetries for leptons
and heavy flavours. The material in this subsection comes entirely from
Ref.~\cite{NRT}.

\subsubsection{$Z^0$ Decay Widths}

Taking into account all radiative corrections, at one loop the partial
width for the decay $Z^0\to f{\bar f}$ is \cite{radcorr}
\beq
\Gamma_{Z\to f{\bar f}} = N_c^f {M_Z\over 12\pi} \sqrt{2} {\widehat G}_\mu
M_Z^2 \rho_f \left( v_f^2 + a_f^2 \right) \left( 1 + \delta_{QED}^f \right)
\left( 1 + \delta_{QCD}^f \right),
\eeq
where $N_c^f = 3(1)$ for quarks (leptons), $\delta_{QCD}^f$ is the QCD
correction for hadronic final states, and $\delta_{QED}^f$ is an additional
photonic correction. Fermion mixing effects appear in two places -- first,
in the vector and axial couplings $v_f$ and $a_f$ (see Eq.~\ref{vecax}),
and also in the effective weak mixing angle which appears in the vector
coupling. This weak mixing angle is renormalized by electroweak effects
\cite{NRT}:
\beq
\label{seff}
s^2_{eff}(f) = {1\over 2} \left[ 1 - \sqrt { 1 -
{ G_\mu \over {\widehat G}_\mu } { 4 {\cal A} \over \rho M_Z^2 }  \left(
{ 1 \over 1 - \Delta\alpha} + \Delta  {\bar r}^{rem} \right) } \, \right].
\eeq
As in the renormalized expression for $M_W$ (Eq.~\ref{wmass}), mixing
effects appear indirectly in $G_\mu/ {\widehat G}_\mu$ (Eq.~\ref{gfermi}).
There are also electroweak corrections in the $\rho_f$ term: $\rho_f = \rho
+ \Delta\rho_f^{rem}$, where $\rho$ contains all large $t$-quark effects
and is universal, and $\Delta\rho_f^{rem}$ (and $\Delta{\bar r}^{rem}$
above) include all the nonuniversal flavour-dependent corrections. In doing
the fits, all corrections have been taken into account, including the
finite mass effects for heavy fermions.

The experimental values of the five partial widths $\Gamma_Z$, $\Gamma_h$,
$\Gamma_e$, $\Gamma_\mu$ and $\Gamma_\tau$ \cite{MZ} have large
correlations among themselves. The widths are all shown in Table 5.

\subsubsection{Leptonic Asymmetries}

On resonance, the forward-backward asymmetry in the process $e^+e^-\to Z^0
\to f{\bar f}$ takes the form
\beq
A_f^{FB} = 3 {v_e a_e \over v_e^2 + a_e^2}{v_f a_f \over v_f^2 + a_f^2}~.
\eeq
As in the partial widths, mixing effects enter both in the vector and axial
couplings (Eq.~\ref{vecax}), and in the renormalized effective weak mixing
angle (Eq.~\ref{seff}). In the fits, all QED and QCD (for hadronic final
states) corrections have been included \cite{NRT}. The $\tau$ polarization
asymmetry has also been measured at LEP. This asymmetry is written
\beq
A_\tau^{pol} = -2 {v_\tau a_\tau \over v_\tau^2 + a_\tau^2}~.
\eeq
The experimental values \cite{asymmetries} for all leptonic asymmetries are
given in Table 5.

\subsubsection{Heavy Flavours}

The partial widths for $Z^0\to b{\bar b}$ \cite{gammab} and $c{\bar c}$
\cite{gammac} have also been measured. These are listed in Table 5. The
forward-backward asymmetries for these final states have also been measured
\cite{FBasym}. For $b{\bar b}$, there is a peculiarity which must be taken
into account. Due to the fact that neutral $B$-mesons can oscillate into
${\overline B}$-mesons, the observed asymmetry is not the true asymmetry
but must be corrected:
\beq
A_b^{FB} = {A_{obs}^{FB}\over 1 - 2\chi_B}~,
\eeq
where $\chi_B$ is a measure of the probability for $B$-${\overline B}$
oscillations. Experimentally, this parameter has been found to be
$\chi_B=0.146\pm 0.016$ \cite{Bmix}. The forward-backward asymmetries for
both $b{\bar b}$ (corrected) and $c{\bar c}$ final states is given in Table
5.

\begin{table}
\hfil
\vbox{\offinterlineskip
\halign{&\vrule#&
   \strut\quad#\hfil\quad\cr
\noalign{\hrule}
height2pt&\omit&&\omit&\cr
& Quantity && Experimental Value &\cr
height2pt&\omit&&\omit&\cr
\noalign{\hrule}
height2pt&\omit&&\omit&\cr
& $\Gamma_Z$ && $2487\pm 10^\dagger$ &\cr
& $\Gamma_h$ && $1739\pm 13^\dagger$ &\cr
& $\Gamma_e$ && $83.2\pm 0.6^\dagger$ &\cr
& $\Gamma_\mu$ && $83.4\pm 0.9^\dagger$ &\cr
& $\Gamma_\tau$ && $82.8\pm 1.1^\dagger$ &\cr
& $A_e^{FB}$ && $-0.019\pm 0.014$ &\cr
& $A_\mu^{FB}$ && $0.0070\pm 0.0079$ &\cr
& $A_\tau^{FB}$ && $0.099\pm 0.096$ &\cr
& $A_\tau^{pol}$ && $-0.121\pm 0.040$ &\cr
& $\Gamma_b$ && $367\pm 19$ &\cr
& $\Gamma_c$ && $299\pm 45$ &\cr
& $A_b^{FB}$ && $0.123\pm 0.024$ &\cr
& $A_c^{FB}$ && $0.064\pm 0.049$ &\cr
height2pt&\omit&&\omit&\cr
\noalign{\hrule}
height2pt&\omit&&\omit&\cr
& $a_b^{\gamma Z}$ && $-0.405\pm 0.095$ &\cr
& $a_c^{\gamma Z}$ && $0.515\pm 0.085$ &\cr
& $A_{c,D^*}^{\gamma Z}$ (29 GeV) && $-0.101\pm 0.027$ &\cr
& $A_{c,D^*}^{\gamma Z}$ (35 GeV) && $-0.161\pm 0.034$ &\cr
height2pt&\omit&&\omit&\cr
\noalign{\hrule}}}
\caption{Partial widths (given in MeV) and asymmetries measured at the $Z$
peak. The correlations among the measurements marked with a $^\dagger$ have
been taken into account in the fits \protect\cite{NRT}. Also displayed are
the axial couplings $a_{b,c}^{\gamma Z}$ and the charm asymmetries with
$D^*$ tagging $A_{c,D^*}^{\gamma Z}$, all measured off resonance.}
\end{table}

Finally, the forward-backward asymmetries for $b{\bar b}$ and $c{\bar c}$
final states have also been measured at lower energies at PEP and PETRA. In
this region, the asymmetries $A_{b,c}^{\gamma Z}$ include interference
between the $\gamma$ and the $Z^0$, and essentially measure the product of
axial couplings $a_e a_{b(c)}$. Both final states are tagged using high $p$
and $p_T$ leptons, leading to large correlations between the two
measurements \cite{ltag}. Due to the correlations, these data are only used
for those fits where only one mixing angle at a time is allowed to vary.
For $c{\bar c}$, there is an additional tagging method not applicable for
$b{\bar b}$, namely using $D^*$'s \cite{Dtag}. The results using this
method are used in all the fits. The experimental data are shown in Table
5. Note that the axial couplings $a_b$ and $a_c$ \cite{axialcoup} are given
in the case of lepton tagging, while for $D^*$ tagging the forward-backward
asymmetry is shown \cite{Dtag}.

\section{Constraints}

In the section, I present the constraints which the experimental data shown
in the previous section place on the mixing between ordinary and exotic
fermions. I will show the results of two fits. In the first (the
`individual fit'), only one mixing angle at a time is allowed to be
nonzero, and in the second (the `joint fit') all mixing angles vary
simultaneously.

In both fits, the constraints are obtained by using a least-squares
method. One complication is that the mixing angles are bounded, that is,
$0 \le s_{L,R}^2 \le 1$. In order to deal with this, the following
procedure is used. For each parameter $s_i^2$, the $\chi^2$ distribution is
calculated. Then, assuming a probability distribution
\beq
P(s_i^2)ds_i^2 = N_i e^{-\chi^2(s_i^2)/2} ds_i^2,
\eeq
in which $N_i^{-1}=\int_0^1 exp(-\chi^2(s_i^2)/2)ds_i^2$ (i.e.~$N_i$ is
chosen such that $P(s_i^2)$ is properly normalized in the domain [0,1]),
the 90\% C.L.\ upper bounds on the $s_i^2$ are calculated from $P(s_i^2)$.

Despite the large number of parameters, the experimental data is
comprehensive enough to constrain all mixing angles. The results of the
individual and joint fits are shown in Table 6, which is taken from
Ref.~\cite{NRT}. In the `Source' column of this Table are listed those
observables which are most important for constraining the mixing angles in
the individual fits. However, in the joint fit it is possible to evade the
bounds from these observables through fine-tuned cancellations between
different mixings. In this case, other observables, which depend on
different combinations of the mixings, become important. These new
observables, which are denoted by a $*$ in Table 6, are typically less
precise, so that the constraints in the joint fit are somewhat weaker than
those in the individual fit.

\begin{table}
\hfil
\vbox{\offinterlineskip
\halign{&\vrule#&
   \strut\quad#\hfil\quad\cr
\noalign{\hrule}
height2pt&\omit&&\omit&&\multispan5 &&\omit&\cr
& \omit && \omit\hidewidth Individual \hidewidth &&
				\multispan5 Joint && Source &\cr
height2pt&\omit&&\omit&&\multispan5 &&\omit&\cr
\noalign{\hrule}
height2pt&\omit&&\omit&&\omit&&\omit&&\omit&&\omit&\cr
& \omit && \omit && $\Lambda=2$ && $\Lambda=0$ && $\Lambda=4$ && \omit &\cr
height2pt&\omit&&\omit&&\omit&&\omit&&\omit&&\omit&\cr
\noalign{\hrule}
height2pt&\omit&&\omit&&\omit&&\omit&&\omit&&\omit&\cr
& $\slesq$ && 0.0047 && 0.015 && 0.0090 && 0.015 &&
			$\Gamma_e,M_W^*,A_\mu^{FB*},eq^*,g_e^*$ &\cr
& $\sresq$ && 0.0062 && 0.010 && 0.0082 && 0.010 &&
			$\Gamma_e,A_e^{FB},A_\mu^{FB*},\nu e^*$ &\cr
& $\slmusq$ && 0.0017 && 0.0094 && 0.0090 && 0.011 &&
		$V_{ui}^2,\nu q,g_\mu,\Gamma_\mu,s_{eff}^{LEP*},M_W^*$ &\cr
& $\srmusq$ && 0.0086 && 0.014 && 0.014 && 0.013 &&
			$\Gamma_\mu,A_\mu^{FB}$ &\cr
& $\sltausq$ && 0.011 && 0.017 && 0.015 && 0.017 &&
			$\Gamma_\tau,A_\tau^{FB},g_\tau,A_\tau^{pol*}$ &\cr
& $\srtausq$ && 0.011 && 0.012 && 0.014 && 0.012 &&
			$\Gamma_\tau,A_\tau^{pol},A_\tau^{FB},g_\tau^*$ &\cr
& $\slusq$ && 0.0045 && 0.019 && 0.015 && 0.019 &&
			$V_{ui}^2,\Gamma_h,\Gamma_Z,eq,\nu q$ &\cr
& $\srusq$ && 0.018 && 0.024 && 0.025 && 0.024 &&
			$\nu q,\Gamma_h,\Gamma_Z,eq$ &\cr
& $\sldsq$ && 0.0046 && 0.019 && 0.016 && 0.019 &&
			$V_{ui}^2,\Gamma_h,\Gamma_Z,\nu q$ &\cr
& $\srdsq$ && 0.020 && 0.030 && 0.028 && 0.029 &&
			$eq,\Gamma_h,\Gamma_Z,\nu q$ &\cr
& $\slssq$ && 0.011 && 0.038 && 0.039 && 0.041 &&
			$\Gamma_h,\Gamma_Z,V_{ui}^2$ &\cr
& $\left(s_R^s\right)^{2\dagger}$ && 0.36 && 0.67 && 0.63 && 0.74 &&
			$\Gamma_h,\Gamma_Z$ &\cr
& $\slcsq$ && 0.013 && 0.040 && 0.042 && 0.042 &&
			$\Gamma_h,\Gamma_Z,\Gamma_c^*,A_c^{\gamma Z*}$ &\cr
& $\srcsq$ && 0.029 && 0.097 && 0.10 && 0.099 &&
		$\Gamma_h,\Gamma_Z,A_c^{\gamma Z*},\Gamma_c^*,A_c^{FB*}$ &\cr
& $\slbsq$ && 0.011 && 0.070 && 0.072 && 0.069 &&
			$\Gamma_h,\Gamma_Z,\Gamma_b,A_b^{FB*}$ &\cr
& $\left(s_R^b\right)^{2\dagger}$ && 0.33 && 0.39 && 0.40 && 0.39 &&
		$\Gamma_b,\Gamma_Z,\Gamma_h,A_b^{\gamma Z},A_b^{FB*}$ &\cr
& $\slnuesq$ && 0.0097 && 0.015 && 0.016 && 0.014 &&
			$s_{eff}^{LEP},g_e,s_{eff}^{NC},M_W^*$ &\cr
& $\slnumusq$ && 0.0019 && 0.015 && 0.0087 && 0.011 &&
			$V_{ui}^2,g_\mu,\nu q,s_{eff}^{LEP},M_W^*$ &\cr
& $\left(s_L^{\nu_\tau}\right)^{2\dagger}$
	&& 0.032 && 0.064 && 0.097 && 0.035 && $\Gamma_Z,g_\tau$ &\cr
& $\sum_{i=4}^n {\widehat V}_{ui}^2$ && 0.0048 && 0.014 && 0.010 && 0.018 &&
			$V_{ui}^2$ &\cr
& $\sum_{i=4}^n {\widehat V}_{ci}^2$ && 0.53 && 0.76 && 0.76 && 0.76 &&
			$V_{ci}^2$ &\cr
& $\vert\kappa_{ud} \vert$ && 0.0011 && 0.0059 && 0.0060 && 0.0058 &&
			$V_{ui}^2,\nu q$,RHC's,$\nu e^*$ &\cr
& $\vert\kappa_{us} \vert$ && 0.0054 && 0.0061 && 0.0061 && 0.0061 &&
			$V_{ui}^2$,RHC's &\cr
height2pt&\omit&&\omit&&\omit&&\omit&&\omit&&\omit&\cr
\noalign{\hrule}}}
\caption{90\% C.L. upper limits on mixing angles for individual fits (one
angle at a time is allowed to vary) and joint fits (all angles allowed to
vary simultaneously) \protect\cite{NRT}. Observables which are most
important for the constraints are shown in the `Source' column (those
quantities which contribute only in the joint fits are tagged with an
asterisk). $s_{eff}^{LEP}$ and $s_{eff}^{NC}$ refer to the weak mixing
angle as extracted in neutral current measurements at the $Z$ peak and at
low energy, respectively. See the text for a discussion of the bounds on
$\srssq$, $\srbsq$ and $\slnutausq$ (marked with a $^\dagger$).}
\end{table}

{}From this Table it is evident that the neutral current data at the $Z$ peak
is especially important for bounding all mixings. For the first generation
fermions and the $\mu$ and $\nu_\mu$, the low-energy charged and neutral
current results (particularly $\nu q$ and $e q$ scattering) are also
useful. In addition, the asymmetries off the $Z$ peak are helpful in
constraining the mixing angles of the $c$- and $b$-quarks.

I must again stress that the data used to obtain these constraints are
already a bit out of date. For example, only the 1990 LEP data was used;
the inclusion of the 1991 LEP data would surely strengthen most of the
bounds somewhat. The most important new development is in $\tau$-decays.
The value of $(g_\tau/g_e)^2$ shown in Table 3 differs from its standard
model value of 1 by about 1.5 standard deviations. However, the latest
measurements of the $\tau$ mass and lifetime have removed this discrepancy
\cite{taunew}. Thus, the limits on $\slnutausq$ shown in Table 6, which
depend on the old value of $(g_\tau/g_e)^2$, should be taken with a grain
of salt -- the new bounds are probably quite a bit better.

In all fits, $\Lambda_e=\Lambda_\mu=\Lambda_\tau$ has been assumed.
Furthermore, in the individual fit, $\Lambda=2$ was taken. Note that, in
this fit, only the neutrino mixings can depend on $\Lambda$. Since
$\slnuesq$ and $\slnumusq$ are bounded mainly by charged current data, the
dependence on $\Lambda$ is minimal. On the other hand, the constraint on
$\slnutausq$ does depend on $\Lambda$: $\slnutausq < 0.098,0.032,0.015$ for
$\Lambda=0,2,4$. (As I said in the previous paragraph, these numbers should
not be taken too seriously. However, even with the new data, the strong
dependence of $\slnutausq$ on $\Lambda$ will persist.)

The constraints on $\srssq$ and $\srbsq$ are considerably weaker than those
of other angles due to a peculiarity of the observables which bound them.
These mixing angles are constrained mainly by the LEP observables, which
depend on the couplings $v_q a_q$ and $v_q^2 + a_q^2$ ($q=s,b$). However,
for $\srqsq\simeq 0.3$, the $s^4$ terms cancel against the $s^2$ terms.
Thus there are two minima in the $\chi^2$ distribution, centred around 0
and 0.3. The 90\% C.L. bounds of Table 6 are obtained by integrating over
both regions. The restriction to the region centred at no mixing gives
stronger bounds, $\srssq\leapprox 0.09$ and $\srbsq\leapprox 0.10$.

The bounds on most mixing angles in Table 6 are quite stringent. However,
one might argue that the exotic fermions which give rise to these mixings
necessarily appear in models with other forms of new physics, extra $Z$'s
for instance, and that these new effects might weaken significantly the
mixing limits. This seems quite unlikely, given the number and variety of
constraints. In fact, such a study has been done \cite{NRTnew}, in the
context of $E_6$ and $SO(10)$ models. In this paper, the effects of
$Z$-$Z'$ mixing and fermion mixing were analyzed simultaneously. In
general, the presence of an extra $Z$ did not much alter the mixing limits.
Although not a proof, this analysis lends support to the idea that,
regardless of the model, it is rather difficult to evade the constraints on
the mixing of ordinary and exotic fermions found in Table 6.

\section{Conclusions}

In this chapter, I have discussed the constraints which precision
measurements put on the mixing of ordinary and exotic fermions. Exotic
fermions are defined as new fermions whose left- or right-handed components
transform in a non-standard way under $\gweak$, that is, $L$ singlets
and/or $R$ doublets. Excluding noncanonical colour and electric charge
assignments, there are 4 types of exotic fermions -- mirror fermions,
vector singlets and doublets, and new Weyl neutrinos.

In general, mixing between ordinary and exotic fermions will lead to
flavour-changing neutral currents among the light particles, which are
extremely well constrained experimentally. However, if one chooses
fine-tuned directions in mixing parameter space such that each ordinary
fermion mixes with its own exotic fermion, then the bounds from FCNC can
be evaded. I have developed the formalism which describes this mixing --
there is one mixing angle per $L$ and $R$ charged fermion. For neutrinos,
the situation is more complicated due to the possibility of Majorana masses
and the fact that there is no experimental evidence against FCNC involving
neutrinos. Nevertheless, because the final neutrinos in any process are
unobserved, it is possible to describe mixing in the neutrino sector by one
angle, plus one auxiliary parameter, per ordinary neutrino species.

There are enough constraints from low-energy charged and neutral current
data, as well as the experimental results from LEP, to constrain all mixing
angles. I have described two types of fits. In the first, all mixing angles
but one are set to zero, and the non-zero angle is constrained. In the
second all angles are allowed to be non-zero simultaneously. The results
are shown in Table 6. In the individual fit, most of the mixing parameters
$\left(s^f\right)^2$ are constrained to be of order 1\%, with some of the
angles (such as those for $e_{L,R}$, $u_L$, $d_L$, $\mu_L$ and
$\nu_{\mu L}$) quite a bit smaller. The two exceptions are $s_R$ and $b_R$,
whose mixings are bounded to be only about 0.3. In the joint fit, due to
the possibility of accidental cancellations among the mixings, the limits
are weakened. Typically, the constraints are relaxed by a factor of 2-3,
but this factor can be as much as 6-8 in a few cases.
\vskip5truemm
\noindent
{\bf Acknowledgements}:

I would like to thank E. Nardi for helpful discussions. This work was
supported in part by the Natural Sciences and Engineering Research Council
of Canada, and by FCAR, Qu\'ebec.


\end{document}